\documentclass[sigconf]{acmart}

\AtBeginDocument{%
  \providecommand\BibTeX{{%
    \normalfont B\kern-0.5em{\scshape i\kern-0.25em b}\kern-0.8em\TeX}}}

\copyrightyear{2025}
\acmYear{2025}
\setcopyright{cc}
\setcctype{by-nc}
\acmConference[CHI '25]{CHI Conference on Human Factors in Computing Systems}{April 26-May 1, 2025}{Yokohama, Japan}
\acmBooktitle{CHI Conference on Human Factors in Computing Systems (CHI '25), April 26-May 1, 2025, Yokohama, Japan}\acmDOI{10.1145/3706598.3713327}
\acmISBN{979-8-4007-1394-1/25/04}

\usepackage{xcolor}
\usepackage{soul}
\usepackage{graphicx}

\begin{document}

\title[Encountering Robotic Art]{Encountering Robotic Art: The Social, Material, and Temporal Processes of Creation with Machines}

\author{Yigang Qin}
\orcid{0000-0001-7843-2266}
\affiliation{
  \institution{Syracuse University}
  \city{Syracuse, NY}
  \country{United States}
  }

\author{Yanheng Li}
\orcid{0000-0002-9767-3468}
\affiliation{
  \institution{City University of Hong Kong}
  \city{Hong Kong}
  \country{Hong Kong}
  }

\author{EunJeong Cheon}
\orcid{0000-0002-0515-6675}
\affiliation{
  \institution{Syracuse University}
  \city{Syracuse, NY}
  \country{United States}
  }

\begin{teaserfigure}
    \centering
    \includegraphics[width=\linewidth]{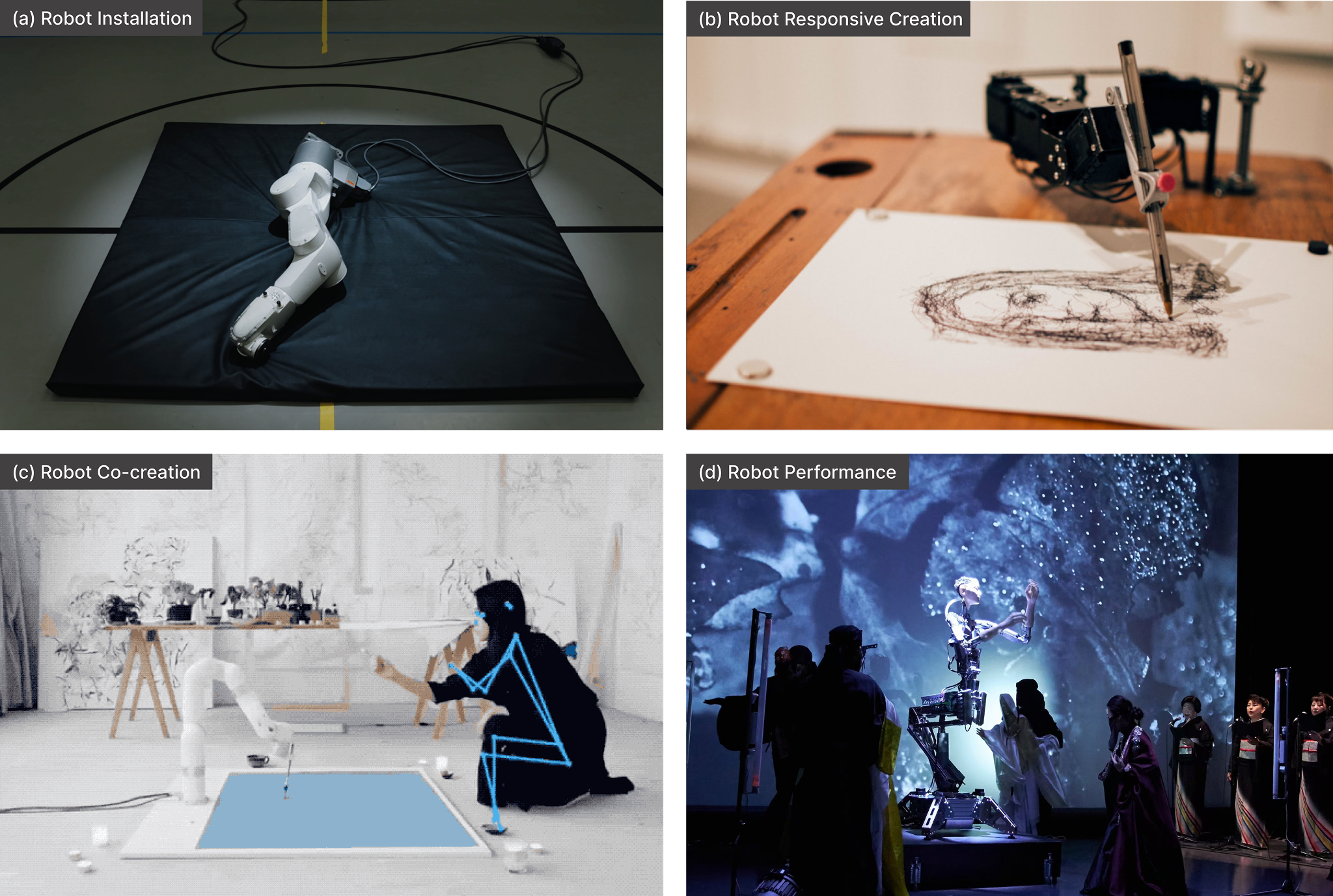}
    \caption[Caption for LOF]{Examples of robotic art: (a) ``\textit{dis}arming II'' by Emanuel Gollob, a robotic installation to explore the relationships between a detached robotic arm, the environment, and the observers~\cite{gollob2023}; (b) ``Human Study'' by Patrick Tresset, a drawing robot with camera depicting live scenes ~\cite{patrick2016}; (c) ``Mutations of Presence (D.O.U.G. 4)'' by Sougwen Chung, a biofeedback-controlled painting robot~\cite{chung2021}; (d) ``Alter 3'' by Takashi Ikegami, a humanoid robot once in performing Kagura, a Japanese dance ritual~\cite{ikegami2021}.}
    \Description{Examples of robotic art}
    \label{fig:teaser}
\end{teaserfigure}

\begin{abstract}

Robots extend beyond the tools of productivity; they also contribute to creativity. While typically defined as utility-driven technologies designed for productive or social settings, the role of robots in creative settings remains underexplored. This paper examines how robots participate in artistic creation. Through semi-structured interviews with robotic artists, we analyze the impact of robots on artistic processes and outcomes. We identify the critical roles of social interaction, material properties, and temporal dynamics in facilitating creativity. Our findings reveal that creativity emerges from the co-constitution of artists, robots, and audiences within spatial-temporal dimensions. Based on these insights, we propose several implications for \textit{socially informed}, \textit{material-attentive}, and \textit{process-oriented} approaches to creation with computing systems. These approaches can inform the domains of HCI, including media and art creation, craft, digital fabrication, and tangible computing.

\end{abstract}



\maketitle

\section{Introduction}

The term ``robot,'' originally signifying `forced labor,' has long been associated with labor and work. Robots have demonstrated their utility in various automated productive and social contexts, where the primary goals are improving productivity, safety, and fostering social interactions with humans~\cite{simoes2022designing, weidemann2021role, honig2018understanding}. However, an increasing number of cases feature using of robots in creative settings. Unlike productive contexts, where the focus is on efficiency and task completion~\cite{arents2022smart}, or social contexts, where communication and trust are prioritized~\cite{nam2020trust, saunderson2019robots}, creative environments prioritize artistic innovation and expression~\cite{hsueh2024counts}. This shift fundamentally alters the dynamics of human-robot interaction, redefining the roles and expectations for both humans and robots.

For instance, robots’ social behaviors are leveraged to support the generation and expression of creative ideas~\cite{hu2021exploring, sandoval2022human, alves2020creativity}, and programmable robotic movements and trajectories are employed to inspire artistic activities such as sketching~\cite{lin2020your}. These studies often engage participants from creative fields who possess limited prior experience with robotics, and are typically conducted in short-term, experimental settings. Consequently, the findings from these studies remain constrained since much can be learned from professional practitioners' experiences to inform system design such as digital fabrication~\cite{hirsch2023nothing}. There is a notable gap in research examining the long-term, active, and practical experience of integrating robotic systems into the creative processes. As a result, the deeper insights into how robots facilitate and shape creative processes, beyond simply augmenting human creativity, remain underexplored. In this study, we aim to better understand the impacts of robots on creative processes and outcomes.

As early as Leonardo da Vinci's 16th century ``Automaton,'' artists have explored the creative affordances of robotic systems~\cite{shanken2002cybernetics, pagliarini2009development, jeon2017robotic}. The artistic creation process typically encompasses various stages, including the exploration of materials and techniques, ongoing experimentation and iteration, and the continual refinement of the artists' insights into their creative subjects~\cite{lewis2023art, sturdee2022state}. Therefore, investigating the artistic process involving robots offers an opportunity to gain deeper insights into robots' creative potential. Robotic art, in particular, provides a compelling case for this exploration.

We define robotic art as artworks that utilize robotic or automated machines to create artistic experiences and tangible artifacts. One example is robotic installation art, in which robots are programmed to follow specific rules that embody the artist’s expression (\autoref{fig:teaser} (a)). Another example is responsive art, in which robots react to their environment, with behaviors that change over time or in response to spectators (\autoref{fig:teaser} (b)). Additionally, there are robotic creators, which possess a degree of agency, allowing them to collaborate with human artists and produce works that extend beyond mere replication of human-created art (\autoref{fig:teaser} (c) and (d)). As such, robotic art becomes a rich case for exploring human-machine interactions in creative contexts. Gaining a deeper understanding of how robots facilitate artistic expression can provide insights for designing computing systems to support creative activities~\cite{gomez2021robot}.

We draw on semi-structured, in-depth interviews with renowned professional robotic artists to investigate the use of robots in artistic practice. Specifically, our goal is to understand how artistic exploration of robotic systems challenges conventional assumptions about the functions of robots, such as their roles in automating repetitive tasks or serving human needs. We also explore the implications of robots in the artistic process and examine how creativity may emerge within robotic art. To address these interrelated inquiries, our study focuses on the practice of robotic art, posing the research question: \textit{How do robotic artists utilize robots in their artistic practice?} We approach this inquiry through the perspectives and experiences of robotic artists, who creatively design, modify, and repurpose robotic systems for artistic expression and exploration.

Our findings highlight the social, material, and temporal dimensions of artists' practices that shape their creativity and artistic outcomes. The creation of robotic art is largely a social process, as artists receive both explicit and implicit feedback through the audience's reactions and reception of their work. Simultaneously, the embodiment and malfunctions inherent to robotic systems drive artistic experimentation. The temporal processes of creation and exhibition, beyond just the final product, further enhance the creative value. Our empirical analysis presents how creativity emerges through the interplay of social, material, and temporal interactions among artists, robots, audiences, and the environment.

We make two main contributions to HCI in this study. 
First, we elucidate the interactive mechanisms among key actors---human creators, machines, audiences, and environments---within the practice of robotic art, a topic that remains underexplored in HCI. Our findings reveal the significance of sociality (e.g., interactions between artists and audiences), materiality (e.g., the embodiment and malfunctions of robots), and temporality (e.g., the processes of creation and exhibition) in shaping creative values. We propose that these three facets are central to the creative process and facilitate the emergence of creativity in robotic art.
Second, drawing from the findings, we offer implications for \textit{socially informed}, \textit{material-attentive}, and \textit{process-oriented} creation with computing systems. We suggest leveraging these three aspects to enhance creativity and the creative experience. Specifically, we discuss the value of incorporating implicit audience feedback, designing with technical malfunctions, and focusing on the post-creation process to foster alternative creative experiences with machines~\cite{alter2010designing, juarez2022glitch}.

\section{Related Work}
As we study practices of artists making robotic art with the intent to inform ways of facilitating creativity, we review literature about robotic art as a case of creative practice to probe the factors that contribute to creativity in the artwork. To justify our contributions as learning from the art to inform design, we review artistic practices in HCI and their unique benefits to the field.

\subsection{Robots in Creative Contexts}

Introducing robots into creative domains provides a new research medium through robots' embodied materiality and interactions \cite{Nitsche2019craft}. Robots applied in creative tasks have been treated not only as executors following a set of predetermined rules and executing commands but also as social actors that can respond to dynamic human behaviors and environments~\cite{nam2023dreams}. For example, robot collaborative drawing highlights robot's role as a tool, to mediate and engage the spatial arrangement, physical activities, and social interactions of people in the co-creation process through robotic motions and creations~\cite{hoggenmueller2020woodie}.  The collaboration between humans and robots in choreography provides a lens to explore how human movement can be reinterpreted and fused by robots, thus inspiring more creative expression~\cite{saviano2024cage, hsueh2019understanding}. In this context, robots function not merely as tools but as partners that contribute to improvisation~\cite{rogel2022music}. Th{\"o}rn et al.~\cite{thorn2020human} examined the potential of robots as dance partners, emphasizing the importance of synchronization of robotic movement and emotional expression in creating a compelling dance experience~\cite{nam2014interactive}. The participation of social robots in storytelling underscored the potential of robots to enhance creative processes through their presence, body language, facial expressions, and vocal modulations~\cite{jeon2017robotic}. \citet{antunes2022inclusive} demonstrated how robots can act as co-narrators or audience members in storytelling sessions, thereby enriching the narrative experience and fostering children's creativity through facilitating multisensory feedback and emotional behaviors.

These examples illuminate the multifaceted roles that robots can play in creative processes, from facilitating creative processes as tools and partners to acting as resources~\cite{chung2021intersection}. To fully grasp the potential of robots in the creative domain, it is essential to explore how their material features and interactive capabilities shape and support creative expression. Beyond the roles that robots play in creative processes, we are also interested in what is unique about robots (e.g., material features and interaction processes) that accounts for their continual advancement and application in creative contexts in a more integrated way. The artistic appropriation of robots may provide part of the answer. Notably, robotic artists engage proactively with robots when creating artwork, exploiting the artistic potential of robots, rather than passively accommodating the robots' inherent limitations. They break down robots, redefine them, and repurpose them as instruments for artistic goals. We propose that robots' specific manifested qualities embed the potential to stimulate creativity and drive such interest and engagement by artists as in their practice. Thus, we aim to better explore the above questions by examining the practices and characteristics of artists' voluntary utilization of robots in achieving artistic ends. Answering these questions can contribute to knowledge about the unique capabilities and roles of robots in creative contexts.

\subsection{Robotic Art}

Robotic art involves the use of robots as both mediums and subjects of artistic expression~\cite{penny2016robotics, kac1997foundation}. Robotic art covers a group of practices that is usually related to computer-automated cultural artifacts (CACA)~\cite{penny2013art}, although some authors refer to the broader field of mechanically driven kinetic art in an attempt to frame it within media art practices~\cite{hoetzlein2009new}. From a general perspective, a collection of works with a certain level of machine autonomy is often addressed as robotic art ~\cite{kac2001origin}, belonging to a broader category of machinic art, which includes `telerobotic art' ~\cite{reichle2009art}, `postbiological life forms' ~\cite{reichle2009art}, `robotic installation art'~\cite{shi2023what}, and so on. Artworks featuring robots not only create more engaging experiences but also challenge the boundaries of intelligent agency and human creativity. For instance, robotic artists prompt us to question whether a robot that creates art can be considered an artist in its own right~\cite{shi2023what, mika2022can}.

Robots have been adopted in artistic expression and creation due to their anthropomorphic qualities, creating a metaphor as social components and providing a degree of sociality~\cite{jeon2017robotic,kahn2014creative}. In this context, artists can leverage robots' social behaviors to stimulate artistic expression. For instance, natural language dialogue can establish social connections with creative producers and guide creative imagination through questioning~\cite{kahn2016human, hu2021exploring}. Artists have utilized robots' facial expressions, body postures, movements, and speech intonation to provide emotional communication during creative theater performances, thereby encouraging the generation of creative human-human/robot interaction~\cite{knight2011eight, li2023nice}. These characteristics distinguish robots from traditional artistic materials such as plaster or fabric, allowing artists to create artwork and express themselves more dynamically through interactions with living materials.

Beyond responding to human social interactions, robotic art attracts artists through its non-static artifact characteristics and adaptive environmental responsiveness, enabling the creation of new forms of interactive artwork~\cite{nelkin1984science}. While traditional art forms are often judged by their fixed final products, much of robotic art's appeal lies in the process—the trajectory of robot movement, reactions to input, and the way outputs evolve over time provide a spectacle that, along with the final result, constitutes the artwork's complete expression. A common example is drawing robots, whose movement trajectories can be identified and preserved by artists as inspiration for artistic imagination or incorporated into the final design~\cite{lin2020your}. By changing drawing tools and materials, robots can transform drawing styles and patterns~\cite{jeon2017robotic, tresset2013portrait, lin2020your}. Mobile drawing robots can also recognize boundaries with creators or the environment to adjust creation trajectories~\cite{moura2007new, lin2020your}.

Thanks to robots' interactivity, robotic artworks often incorporate audiences into the creation process, allowing artists to harness more collective creativity and enabling greater possibilities in artistic expression~\cite{vavara2022, mikalauskas2018improvising}. Examples include Tresset's performance installations with \textit{Paul}, a custom robot art and computer program for real-time portrait drawing~\cite{tresset2014artistically}. In these works, audiences not only observe a robot creating art but are invited to witness the creative process of a technological ensemble at work, observing how human artists and robotic systems continuously influence each other~\cite{chung2022sketching, gomez2021robot, tresset2014artistically}.

While these researches have demonstrated the potential of robots in artistic creation and their capacity to stimulate creativity, they discuss discretely how creativity is supported and unleashed by single robotic systems, revealing limited approaches toward creativity. There remains a need for research that explores multiple cases involving diverse forms of robots to deepen our understanding of how robots can contribute meaningfully to the creative process. Our study addresses this gap by analyzing the creative practices of multiple robotic art practitioners working with various mediums, such as drawing, dance, and installation. This approach allows us to uncover shared experiences and insights across different artistic forms, potentially revealing broader principles for how robots can effectively facilitate and shape creative practices.

\subsection{Artistic Practices in HCI}

The relationship between art and technology is becoming increasingly intertwined, with artistic practices being extensively integrated into the HCI community to study human interactions with machines in specific contexts~\cite{jeon2019rituals}. By adopting exploratory approaches, these practices challenge traditional usability-study research design. For instance, some studies involve participants in spatial interactive sound installations and use sound theory and micro-phenomenological interviews to analyze the interplay between sensory perception, bodily intent, and sound~\cite{Robson2024sound, Frid2019sound} capturing the subtleties of how users perceive and engage with sound in immersive environments that are often overlooked by conventional measurement techniques. ~\citet{Rajcic2020mirror} introduced an emotion-responsive mirror that generates poetry in real-time, exploring the contrasts in emotional cognition, processing, and expression between machines and users. Similarly, \textit{Being the Machine} by \citet{devendorf2015being} invites artists to perform 3D printing manually in a machine-like manner, probing questions of agency and control in human-machine interactions. Furthermore, collaborations with choreographers have led to the creation of movement-sonification scarves as probes in dance performances, investigating dancers' engagement with music and aiding HCI researchers in exploring how to design body-centric interaction technologies~\cite{Bang2023dance}. The creation of artistic works, with their inherent ambiguity and potential for unforeseen human interactions, parallels the challenges in studying human interactions with emerging technologies~\cite{coessens2009artistic}. Interactive artworks empower HCI researchers to expand their understanding of emergent human behaviors in relation to machines.

However, the role of artistic practice in HCI research has been a subject of ongoing debate with some researchers questioning integrating artistic knowledge into HCI~\cite{frankjaer2020knowledge, candy2011research}. Our review suggests that examining the relationship between art and technology serves as both inspiration and a means to understand the interaction and the broader social world~\cite{taylor2017performing}, leading to novel design research methods such as cultural probes~\cite{gorichanaz2020public,gaver1999probe}, participatory/co-design~\cite{clark2016situated, disalvo2007mapover,soden2020disaster}, and arts-based ethnography~\cite{kang2022electronicists}. Some studies incorporate interdisciplinary teams in the creation and performance of artworks to explore nonlinear collaboration across disciplines, highlighting the intersection of technological and aesthetic thinking
~\cite{kang2022electronicists,kang2018intermodulation}. Research on self-situated performance practices expands our understanding of new relationships and interactions between researchers and participants in HCI, allowing for ongoing refinement of research practices~\cite{taylor2017performing}. In these cases, artistic practice is not merely an object for investigation but something that implies broader issues of research, offering art creators, participants, and researchers opportunities for reflection and knowledge transformation.

Besides, artistic practices also bring diverse aesthetic experiences to the HCI community, helping researchers focus more on the sensory dimensions of interactions with technology. The concept of aesthetic computing, exploring how art and aesthetics are applied in different areas of computer science, introduces a sensuous aesthetic framework to the exploration of technology in HCI research~\cite{fishwick2006aesthetic, jeon2017robotic}. Artistic practices bring embodiment and interaction to the forefront, placing human sensory experience, judgment, and contemplation at the core of aesthetic experiences~\cite{taylor2017performing}. For example, soma design focuses on elevating the sensory experiences of bodily activities to consciousness, utilizing these experiences as resources in the design process. Designers engage deeply with materials—whether physical, digital, cultural, or movement-based—through all their senses, cultivating their aesthetic sensibilities to create innovative designs~\cite{hook2018designing, hook2016soma, luft2023boards}. Consequently, various artistic practices, including choreography~\cite{allen2022choreo,eriksson2019drone}, sound art~\cite{kilic2023corsetto}, and media art~\cite{branch2021tele}, are increasingly incorporated in HCI to shift the focus to bodily sensations, prompting a reevaluation of our judgments about technology.

Integrating artistic practices into HCI research has shown significant potential in broadening our understanding of user experience and technology design. This study explores how robotic art can help the HCI community deepen its understanding of the relationship between human creative behaviors and robots as facilitative materials. We aim to bridge the gap between artistic practice and technology design for creativity by proposing a framework explicating how computing systems like robots serve as the tools and sites for creative endeavors and offering subsequent design implications for creative activities.

\section{Method}

\subsection{Participant Recruitment}

In this study, we conducted in-depth, semi-structured interviews with nine artists working on robotic art. They are renowned professional artists who use or build physical robots for their artistic work, namely robotic art. Designers and amateurs are out of this criterion. We identified potential participants by browsing individual and artistic organizations' websites during the initial recruitment. Subsequently, we employed snowball sampling to recruit more artists who met our criteria. The recruitment ended when iterative data analysis reached saturation. The vast majority of the participants have had formal art training in drawing, painting, installation, media art, music, sculpture, or dance, and all of them incorporated robots into their particular genres. Some have technical backgrounds; many have a doctoral degree in the fields of art or technology. All of them build their robots or at least modify out-of-shell robots. \autoref{tab:demographics} shows more demographic information. 
The study is approved by the Institutional Review Board of the lead author's institution. The artists provided oral consent to participate in this study voluntarily.

\begin{table*}[htbp]
\centering
\begin{tabular}{c c c c c c}
\toprule
Participant & Gender & Age & Geographical Base & Occupation & Years as Artist \\
\midrule
Alex & Man & 30+ & East Asia & University \& art & 10+ \\
Daniel & Man & 30+ & Australia & University \& art & 10+ \\
David & Man & 50+ & Europe & Full-time art & 30+ \\
Evelyn & Woman & 50+ & North America & Full-time art & 10+ \\
Linda & Woman & 30+ & North America & Non-profit \& art & 10+ \\
Mark & Man & 40+ & Europe & Full-time art & 10+ \\
Robert & Man & 60+ & North America & Full-time art & 30+ \\
Samuel & Man & 60+ & East Asia & University \& art & 20+ \\
Sophie & Woman & 30+ & Europe & Full-time art & 10+ \\
\bottomrule
\end{tabular}
\caption{Participant demographics in alphabetical order of the pseudonyms.}
\label{tab:demographics}
\end{table*}

\subsection{Semi-Structured Interviews}

Before the interviews, the first author reviewed the media press on the artists' websites, gaining background information to build rapport, develop interview questions, and inform analysis. He conducted the interviews via Zoom with video recording, which lasted an average of 67 minutes. The in-depth interviews enabled us to look at the layers of meaning that artists relate with their work---unraveling feelings, attitudes, values, and motivations crucial for addressing our research question~\cite{pugh2013good}. The initial interview questions were structured around the following topics: the workflow, motivation, and rewards and challenges of working on robotic art; the relationship between artists and robots; and the source of creativity. These inquiries were driven by our curiosity in making sense of the artists’ work practices. We aimed to understand their motivations for engaging with robotic art, unpacking the values embedded in their practices and informing broader topics around human-machine creation. 

As the interviews began, the first author asked the artists to \textit{introduce themselves and their artwork}. Then he asked about \textit{their inspiration source for their artwork and motivations for working on robotic art}, steering the conversation toward creativity and the characteristics of robotic art. Following that, he solicited their reflections and anecdotes on the \textit{opportunities and challenges of working on robotic art} to understand the nuances of robotic art practices. He then guided the conversation to explore the \textit{relationships between artists and robots}, focusing on their respective agencies and contributions to creative outcomes in art practices. Additionally, he posed questions regarding the artistic practices' \textit{embodied/material engagement}. Concluding the session, he asked the artists to elaborate on any aspects they wished to add. To comply with the semi-structured nature, he prompted follow-up questions throughout the interviews and remained open to new ideas introduced by the artists, allowing some emergent ideas to lead the conversation. After each interview, the artist received a \$40 Amazon or Visa e-gift card if they chose to accept compensation.

\subsection{Qualitative Analysis}

Data collection and analysis proceeded in parallel and iterative ways throughout the study. We began the analysis shortly after conducting each interview. Emerging insights from the analysis led us to refine our interview protocol several times, focusing on the most salient story in the data. The analysis also informed theoretical sampling~\cite{charmaz2006constructing} at the later stage, purposefully recruiting participants who are more likely to provide missing components or revise the emergent themes to construct robust findings.

The analysis mainly followed Charmaz's constructivist grounded theory approach~\cite{charmaz2006constructing} and used Atlas.ti. The first author began with line-by-line open coding on the interview transcripts, followed by focused coding to synthesize and organize the initial open codes into categories. He compiled analytical memos after each coding phase to document emerging categories, reflect on assumptions, and enable constant comparisons across participants, codes, and categories. The process was iterative and collaborative with other authors, leading to 101 codes in 9 categories. For example, the code \textit{VALUE: process} within the \textit{VALUE} category marks discussions around artistic values inherent in creating and presenting robotic art. The code \textit{RELATIONSHIP: instrument} within the \textit{RELATIONSHIP} category illustrates how the artists perceive robots as instruments for their artistic pursuit. The numbers of codes and definitions of all categories are listed in \autoref{app: coding}. Subsequently, we assigned high-level theoretical codes to the categories to delineate the relationships among the categories. This theoretical coding resulted in four themes explaining the generation of creative values in robotic art practice: \textit{Embodied and material presence}, \textit{Technological malfunction}, \textit{Audience reception and feedback}, and \textit{Creation and exhibition processes}. The iterative process forged us to construct a coherent narrative that well-represents the data. The analysis reasonably reached saturation as participants recruited later to a large extent repeated previous artists when answering questions regarding our key themes.

When the analysis was complete, we conducted a member check---shared the Findings section with the artists, asked for comments, and revised the findings when appropriate. Nevertheless, our positionalities inevitably shaped the analysis. The first and third authors have limited experience in professional art practices, which might constrain their ability to fully grasp the nuances of the artistic process. The second author, with an educational background in design and arts and publications in art exhibitions, brings complementary artistic expertise to the analysis.
\section{Findings}

The inductive analysis across different robotic artists revealed recurrent factors that contribute to artistic creativity in robotic artwork. Here we present four such facets---\textit{Embodiment and Materiality}, \textit{Malfunction}, \textit{Audience's Reaction and Reception}, and \textit{Process of Creation and Exhibition}. Robotic art is unique in each of them. We argue that these factors are salient in the real-world practices of robotic art---uses of robots in artistic or creation activities. By investigating the practice of robotic art, our study contributes empirically to understanding how computing machines are creatively used for artistic and non-pragmatic purposes. Building upon prior works on artistic input to HCI ~\cite{kang2022electronicists}, we advance the discourse by exploring how artistic practices, values, attitudes, and ways of thinking can serve as resources for HCI practitioners studying or designing for creative activities with machines.

\subsection{Embodiment and Materiality}
\label{f:emb}
Embodiment and materiality are key factors in artistic creativity, shaping the design of robotic artworks. As embodied forms, robots interact with physical space, materials, and humans, matching with human cognition through bodily perception~\cite{davis2012embodied}. Their embodiment encompasses physical appearance, movement, and human interaction, aspects crucial for HCI researchers designing robots to engage with their environment~\cite{marshall2013introduction}. For most of our artists (N=7), understanding robots' material and embodied nature deeply influences their creative process, shaping their thinking and inspiring new ideas. While embodiment imposes physical limitations, it also enhances artistic expression, fostering new styles and aesthetics.

\paragraph{Expressivity From Embodiment}
The embodied property of robots produces an important expressivity and artistic style in robotic art that is challenging to replicate without physical embodiment. For example, David compared drawing by physical robots with drawing in computer programs, concluding that the former is more expressive in an artistic sense because the action of drawing by robots is embodied in the physical world rather than being ``simulated'' in computer programs: ``\textit{I use embodiment (embodied action of drawing by robots)... the drawings work because they do real gestures, it (the drawing) is not simulated. So the drawing has this dynamic feel to it because it is really the movements and the gestures and things... there is a certain speed that it (the embodiment) gives this expressivity to the drawing}.'' The embodied drawing by robots adheres to the physical properties of the material and environmental factors (e.g., pencil, paper, table, robotic arm's degree of freedom, humidity, lighting of the scene), making the drawing process complex, and at times, random and uncontrollable. This complexity introduces more possibilities for artistic expression.

The degree of artistic expressivity depends on which specific materials enable the embodiment of drawing by robots. Interestingly, David claimed that industrial robots, though can draw with high precision, produce less expressive drawings than his self-built robots whose robotic arm's movement is not that precise but more dynamic, flexible, and turbulent:

\begin{quote}
    I don't use industrial robots, because industrial robots are pen plotters. They do exactly what you ask. But they (non-industrial robots) are flexible and... not that precise... when it's drawn, it (the drawing by non-industrial robots) has more expressivity because of the embodiment. The embodiment is very important. It's only because I use those types of arms (self-built robotic arms). It would be far less important if I was using industrial robots.
\end{quote}

He also mentioned explicitly that precise drawing is not artistic: ``\textit{But anyway, that (precise drawing robot) is the technology. And it works very nicely, but I couldn't find it artistic. I was actually disappointed when I got it to work.}'' Similarly, Sophie noted that plotting/printing robots create different drawings than painting robots do: ``\textit{I wanted it (the artwork) to be painted and I didn't want it for the visuality of it or the behavior of it. I didn't want it to be plotted or printed, [it] feels different [and] has a different existence.}''

Although both industrial robots and self-built robots draw in embodied ways, the results can appear either precise or dynamic, depending on how the robots are built and programmed---in other words, how the artists configure the material aspects of robots to realize the embodiment. In practice, our robotic artists need to think about ways of utilizing embodiment and properties of robots and all other involved materials to be artistically expressive, to be creative.

\paragraph{Inspiration From Embodiment}
We found that the embodied nature of robotic art often becomes a source of inspiration for new artistic ideas. Linda, an artist-engineer who conducts scholarly research at the intersection of robotics and dance, reflected on how interacting with embodied robots makes her think about the differences between human and robotic bodies:

\begin{quote}
    I've never felt more human. You just feel, you notice, oh, I can fall here and I can get right back up, but it (the robot) falls and it can't get right back up. Or how soft am I? How wet? Like, (patted her face) there's so much water content and squish when I lay on the floor. And it [the robot] doesn't have that... That generates new ideas and helps me be creative.
\end{quote}

She also explained how robotic bodies allow her to examine human movements: ``\textit{The robot is doing something that I can't do on my own body---pure right (her arm was moving toward her right), and [then]... [I] can look at my messy right [movement] next to its [robot's] pure right... that's creative, that's energizing to me to see and play with movement profiles with such a pure tool for decomposing the elements of it, making me notice them}.'' She also shared an anecdote that building a special robot with high degrees of freedom inspired her to explore the differences between human and robotic bodies, enabling her to see new things for her art projects.

\paragraph{Creativity From Embodiment}
Our artists emphasized the embodied nature of creativity and intelligence in general based on their artistic practices, asserting that creativity is inherently embodied rather than disembodied, symbolic, spiritual, and something only happens in the human head. For our robotic artists, creativity is built upon understanding embodied entities in the environment rather than abstract concepts in the mind. Samuel used ChatGPT as an example to argue how the disembodied way of communication between humans and machines limits creative interactions:

\begin{quote}
  I think most of the creativity is coming from non-verbal information flow. So when we are discussing with ChatGPT only through text... the creativity that we can experience is so limited because we do have to sit in front of ChatGPT and we cannot move around or ChatGPT is not going to move around. So our conversation is... very limited... that missing embodiment... is also missing creativity in the conversation with ChatGPT.
\end{quote}

The design of ChatGPT aligns with the mainstream approach to disembodied chatbots running as computer programs. In these designs, symbolic content (e.g., text, images, videos, audio) serves as the communicative medium, but bodily interaction is minimized (i.e., users primarily sit and type). While creativity is arguably rooted in embodied interaction with other material bodies, current interactive agents (e.g., Copilot and Midjourney) designed to support creative work remain largely symbolic and disembodied. Limiting human-machine communication to symbolic channels may lose the benefits of embodiment in acquiring creativity.

According to our artists, one reason for the lack of attention to the embodied dimension of creativity is the historical dichotomy between mind and body, which categorizes creativity as something in the mind:

\begin{quote}
     It (the idea that creativity is disembodied) was very much driven by a view that you can split the body and the mind, and intelligence is happening in a symbolic way, mainly in the brain... [This] led to a large focus on software applications and delayed focus on robotic hardware improvements. And still today, you can see the split of hardware and software... [F]or a lot of organic entities, the integration of bodily capacities with their environment could be seen as more intelligent than the representational capacities... [A]s an artist, I am trained to work with bodies interacting with environments or with other bodies, also this fluent transition from bodily action to semantic questioning. (Daniel)
\end{quote}

This dichotomy, which may have formulated the engineering of computing systems, is rarely compatible with the artist's view that intelligence and creativity can be more richly manifested through bodily interaction and relationships.

Embodiment has been an essential prerequisite of creativity for some artists since their creative production requires understanding embodied entities. To summarize this subsection, embodiment is an important source of creativity for robotic artists. Practically, it yields new artistic expressions and aesthetics whose complexity is difficult to replicate by computer programs. The embodied form of robots, in turn,  inspires creative ideas for artworks. These ideas can arise from understanding the entities embodied in the physical world, whether robots, humans, or other bodies in the environment. The symbolic and disembodied modes of interaction between human creators and machines in creative activities can be complemented and strengthened by embodied interaction.

\subsection{Malfunction: ``Ghost in the Machine''}
Robots, encompassing both mechanical and digital devices, are inherently susceptible to malfunction, with physical robots being more prone to errors, glitches, and noise than virtual agents. These malfunctions are widespread in robotics. In robotic art, such errors hold unique significance, influencing the interpretation and value of the art. Unlike engineers, who aim to fix errors, robotic artists often embrace malfunctions as part of their creative process (N=7).

\paragraph{Embracing Errors and Uncertainty}
Evelyn views machine errors not as obstacles, but as opportunities for unique artistic expression. She embraces the imperfections that arise from machine errors, seeing them as a way to humanize the machine and its output:

\begin{quote}
     I embrace these errors. For me, it is the way to show that using the machine in a way that's very counter-intuitive... celebrating that error instead of trying to perfect it, or slowing down the machine instead of trying to create commodities as fast as we can... what's interesting with the machine [is] to actually turn it upside down and think that the machine is a bit like a human child, and everything it does actually slow, it's imperfect, it's full of mistakes.
\end{quote}

Evelyn’s approach challenges the conventional expectation of machines as flawless and efficient executors. By slowing down the machine and celebrating its errors, she imbues the machine with a human-like quality of imperfection. This perspective turns the machine into something capable of growth and learning, much like a human child. The errors, therefore, potentially become a source of uniqueness and individuality in the artwork, adding depth and complexity to the artistic expression. This ``counterintuitive'' way of viewing error resonates with Alex who contrasts this view with the engineering tendency that strives to be neat, rational, and organized through monitoring and fixing errors: ``\textit{[S]ometimes, it's (error/glitch is) like a source of treasure. Like you find something that you could work on, you find something that people don't really use... But when we are tinkering, we sometimes reach this point of, `ah, okay, now this is visible.'... I think sometimes even just those things (errors/glitches) could be a work of [art]}.''

Alex and others see robotic malfunctions as opportunities to imagine alternative approaches and values. When robotic systems' behaviors deviate from their programs, they often refuse to ``fix'' the unexpected behaviors, instead, they allow the unexpected to unfold as serendipitous events that can inspire new design features. Preserving malfunctions allows the artists to think about the artistic potential of something derailing from the initial plan and make informed adjustments accordingly. These values would not be examined, integrated, or utilized to contribute to creativity if the immediate response to malfunctions was negation and subjecting the malfunctions as inferior to the planned behaviors. As Sophie noted, artistic practices are inherently unpredictable and shaped by the contingencies of the creation process.

Many of our artists described how they perceive, evaluate, and appreciate the unexpectedness of robotic art, revealing new artistic ideas that would not have emerged otherwise: \textit{``[I]nstead of an ink particle, you had a hole in the form of that part... I was like, `Oh, we'll see that the material is saturated, I will not push it (brush).' But the robot doesn't have this understanding and pushes it. And I thought, `Oh, it's actually a good outcome. It's actually both conceptually and aesthetically very pleasing to me'}'' (Sophie).

In this case, the robot performed an action that a human artist normally would not perform---pushing the brush on the canvas. The robot made an unusual decision and breakthrough in expression, called by many artists as ``surprise.'' Once the artist recognizes its artistic value, it may be further explored and developed. Linda shared a similar anecdote where an unexpected jitter from the way the motor pulls a string gives a ``texture'' to the robotic movement, which she sees as creative.

\paragraph{Incorporating Malfunctions as Intended Design}
Our artists deliberately incorporate errors into their artworks. It demonstrates how valuing malfunctions and the unexpected can directly contribute to the work's artistic creativity. Linda articulated the idea that humans are capable not only of learning from mistakes but also of intentionally leveraging these errors to their advantage, echoing insights from our other artists: \textit{``Glitches are 100\% part of the creative and artistic process... It's undeniable that we recover better from mistakes [than computers do], but I think it's more than that. We actually can incorporate mistakes and make them part of an intended design.''}

By making malfunctions part of the intended design, the artists engage with and utilize them to enhance artistic expression or similar ends. Choosing not to fix these issues offers the artists alternatives to designing and realizing their robotic art. For example, David recounted an anecdote about a bug---a flaw in a computer program's software or hardware---that unexpectedly made a line drawing ``beautiful.'' Rather than fixing the bug, he decided to make it an optional feature, allowing him to switch it on or off:

\begin{quote}
    Generally I don't take care of them (glitches). So there are those glitches that give this unpredictable because I like to have drawings that are not predictable... I fix it (bug) and then I have the possibility of using or not using the bug... I'm always surprised by the output... it (bug) creates a surprise for the spectator who is looking at the robot drawing... I just left it (bug) and it's still there. Sometimes I switch it (bug) on, sometimes I switch it off.
\end{quote}

If David had fixed the bug, without retaining it in the program, he would not have possessed such a feature of expression. This shows how differently robotic artists handle technical malfunctions than typical engineers or roboticists. Malfunctions should be avoided in engineering but may yield creative outcomes for robotic art. This is not to claim that the creative value is innate within malfunctions. As our findings have shown, malfunctions are raw materials that can be deliberately utilized by the artists to achieve creativity. When malfunctions are not desirable in art, they may primarily be engineering challenges, as the following examples illustrate.

\paragraph{Avoiding Malfunctions}
The fragility of robots is a widely shared concern among our robotic artists (N=8). Regardless of their origin---self-built, modified, or off-the-shelf---all robots are susceptible to breakdowns in real-world environments, particularly during extended exhibitions without proper maintenance by artists or qualified personnel. For exhibitions, malfunctions are generally unacceptable, and robots ought to \textit{perform flawlessly} when showcasing to the audience. To address malfunctions, the artists came up with different strategies, such as having backup materials for replacement and assembling the robots on-site at the exhibition. One strategy is reducing the complexity of the robotic system, simplifying it to minimize the risk of failure or loss of control. Their approach involves designing robots that resist internal breakdowns and withstand external environmental factors, such as moisture and gravity. Linda explained, ``\textit{If I do build them (robots), I try to keep them simple and I try to make something that will withstand its environment... Sometimes that might be outdoors next to the ocean for six days}.'' David further emphasized that the concern for fragility leads to the need for simplicity in robotic design:

\begin{quote}
    [I]f you're used to do programs that are disembodied, that are only on the computer, you can do very complex things. But as soon as you work with robots, you have to simplify everything... They exist in the same physical world [as us]. Dynamic, the speed, the time, the weight of thing are the same for us. So there are all those limits, which [requires] you to simplify a lot of the programming.
\end{quote}

Mitigating malfunctions and recognizing their artistic potential are not mutually exclusive. Designs that address engineering malfunctions can also yield artistic qualities. As illustrated in the findings, utilizing and mitigating malfunctions occur at different phases of artistic practice. In the production phase within the studios, artists often regard malfunctions not as impediments but as sources of inspiration. By celebrating serendipitous errors and the unexpected, they deliberately integrate these elements into their robotic creations, pushing material and expressive boundaries. In this phase, the primary interaction happens at the individual level---between the artist and the robot(s). In contrast, within exhibition spaces like museums, malfunctions conflict with the expectation that the robots should function flawlessly, risking being disqualified from display. Here, the interaction shifts to a social context, where artists must negotiate with curators and audiences on how to present the robot. This transition from studio to exhibition thus signifies an important change in, context, practice, and actors involved. Hence, next, we highlight the significance of audience reaction and reception that shape creative outcomes.

\subsection{Audience's Reaction and Reception}

\begin{quote}
    I suppose [that] every project I do is a collaboration between me, the machine, and the interactant to some extent. --- Robert
\end{quote}

The artistic and creative value in robotic artwork is determined not just by the work itself but often by the audience’s reactions and interpretations. Our artists (N=7) mentioned that they observe or think about audience reaction, and often incorporate them into subsequent iterations of their work. Alex, for instance, is motivated in the first place by observing how people react to robots, drawing inspiration from their perceptions.

\paragraph{Audience Reaction Shapes Robotic Design}
One of the most direct ways audiences influence the practice of robotic art is through the artists, even when it is unintentional. For instance, after observing that some audience members interact with his robots by squeezing two springs on the robot together---causing a short circuit---Robert decided to revise the material design of the robots to prevent such accidents: ``\textit{I knew darn well that the children were going to squeeze the springs together. So I was very excited to find that even if they did that, I put a kind of a self-healing fuse, polycrystalline that will heal itself... it was an important component of the design}.''

Robert’s response highlights the importance of audience reaction, which he observes and integrates into his robot designs. While in this case, the reaction led to the resolution of a technical issue rather than adding an artistic element,  Alex's experience illustrates how audience interaction can inspire new aesthetics in his work. Alex described how he adapted the environment around his robots based on the audience’s tendency to project personalities onto them:

\begin{quote}
    People project something like animals or themselves or something [on the robots]. And then I got inspiration from that. Then I made a little brighter setup with some objects, a little bit like forest kind of setup. And then people try to imagine more stories. And then I also put some effect to [make the setting] looks like night or daylight or morning. Then people really see [the robots] differently.
\end{quote}

These examples demonstrate how the audience's explicit and implicit feedback (action, projection, and imagination) influences artists’ decisions in designing robots. Audiences are not passive recipients of the artists’ creations; rather, they become part of a collective creative process, leaving their mark on the final work.

\paragraph{Audience Reaction Shapes Robotic Performance}
Linda described how she designed a robotic component for ``\textit{onstage performer[s] as well as audience members to come and interact with [the] robot in a creative way},'' emphasizing the importance of creating a space for audience interaction. Robert further suggested that these interactions during the exhibition possess performative features, which he views as an artwork: ``\textit{I would consider the final product (the drawing by his autonomous robots) as the art. And I would also consider the [audience's] experience of watching them (the robots) paint also as a kind of performative artwork}.'' Robert views robots not as static objects but as responsive entities capable of meaningful interactions with both their environment and the audience. He views robots as possessing ``emergent agency'':

\begin{quote}
    I think that's an agency I would call emergent agency, which is to say that the system software, the physical structure itself in relation to the viewer, interactant creates a kind of emergent behavior where the robot is, and it's designed to some extent to react or respond either with sound or motion in some way to the viewer. And by doing so, it then allows the viewer to see that response, which then reprograms the viewer's response to that. So there's almost a kind of feedback loop that I find happens a lot with robotic art.
\end{quote}

Daniel mentioned a similar idea in the context of live dance performance. The performance benefits from incorporating ``real-time learning interactive systems'' because that makes the performance not solely predefined but ``\textit{[emerged] in the moment of interaction which was not there before [the performance].}'' Without the audience serving as the stimulus, interactive robots in exhibitions would not be perceived as they were. In other words, robots react to the audience, which casts changes in the audience's perception, then robots sense the changes and react again, forming a continuous feedback loop or improvisation between the robots and the audience.

\paragraph{Open Interpretations Make Robotic Art}
Artwork that remains open, undetermined, complex, and vague often invites diverse interpretations~\cite{eco1989open}. The same applies to interaction design where systems may not have a single user interpretation~\cite{sengers2006staying}. Samuel built three humanoid robots with different levels of functionality. The third robot, though technologically more advanced, received less curiosity from the audience than the first, more rudimentary robot:

\begin{quote}
    [For the third-gen robot],... people immediately understand what he (the robot) is doing. So people just leave after five minutes. But [for] the first one (first-gen robot), people tend to spend like 20 [or] 30 minutes because people don't understand what he's doing. But now it [the third-gen robot] is interpretable, so I understand that... giving him too much meaning is dangerous, [when] work[ing] on an art stuff, because people get tired... people are used to those things (technological functions), which [have] tons of meaning [about] what the machine is doing.
\end{quote}

Here, incorporating technical functionalities into the robot assigns clear objectives easily grasped by the audience, making the perceived meanings more rigid and restricting the scope for diversified interpretation.

Beyond the individual level, the way of interpretation is also socially shaped. Samuel made the point that the perception of creativity is also partly a social product because ``\textit{creativity is depending on what kind of society we are in and what kind of people we are interacting with}.'' Mark and Robert extended that the perception of robotic art is culturally conditioned, varying across different societies and generations. They mentioned how the animist cultural tendency of some East Asian societies potentially makes people more willing to accept and interested in robots and non-human entities (e.g., plants and animals) behaving as if intelligent and agentic. The way that the social context of interpretation and perception determines artistic values reiterates our claim that the audience's reception of robotic artwork is one of the key aspects of robotic art practice. It suggests that in achieving certain artistic goals by robotic art, considering the audience's background and ``horizon of expectations''~\cite{jauss1982toward}---the socially and historically conditioned structure by which a person comprehends, interprets, and appraises any text based on cultural codes and lived experiences---may be constructive in refining the work's idea.

\subsection{Process of Creation and Exhibition}

Many of the artists we interviewed (N=6) emphasized, or alluded to, the artistic value in the \textit{process} of making robotic art. Specifically, two types of processes are discussed here---the process of \textit{creation} and the process of \textit{exhibition}, reflecting two salient temporal stages of robotic art practice. We do not, by any means, suggest that process is unique to robotic art; apparently, other forms of art also attend to processes of their art practice. Our intention has been to situate the analysis of process in the emerging, particular context of robotic art and to reveal how process leads to a new understanding of robot's uses and roles in real-world scenarios.

Sophie builds robotic systems capable of physically painting on canvas. She uses these robots to explore the painting process itself rather than to focus on the final product—what she referred to as images instead of paintings. Her case exemplifies that the act of making can become the focal point of artistic interest. In her view, paintings as artifacts are space-and-time bound ``material-based work'' that requires ``interactive practice'' and ``decision making,'' whereas the resulting images are ``merely digital representation[s]'' of this process. The difference between images generated by computer programs and paintings created through human touch underscores her rationale for utilizing robots: to bring the tactile, material process of painting to the forefront.

\begin{quote}
   [I]n the end, if I'm trying to crop everything (all my ideas) together, then it (the commonality) is to make the temporality of the decision making process of painting more visible and present. So I'm not really interested in how the image looks. And we experience an object that actually has a temporal element, how it's been created with layers, with tons of decision making... because I am interested in painting as a process and less [as] a product, I'm trying to use the process of making a painting to reflect a lot of our human creativity, our relationship to machines, questions of agencies, and so on.
\end{quote}

She has been building robotic systems that have ``adaptive behavior[s]'' during the painting process, where the systems are designed to ``\textit{analyze a stroke [on the canvas] and then create a successive one}.'' This design ensures that robots' actions are not exclusively dictated by the pre-programmed instructions but also influenced by the constantly changing ``state of the world,'' which includes factors such as the evolving canvas, environmental conditions, and the interaction between the robot and its surroundings. Consequently, a painting is not just a visual product but represents a series of actions with a temporal dimension.

Another important process for robotic art practice is exhibition. In the exhibition space, robotic artworks often take the forms of performances or improvisations, actively interacting and potentially shaping their environment in real time. 
For example, Alex's robots paint spontaneous color patterns on canvas during the exhibition, transforming the event into a performative art experience that aligns with his intention of foregrounding the painting process. The dynamic nature of live drawing at the exhibition---``making a show live''---has been central to Alex's artistic approach.
Moreover, new qualities of robotic artwork not only emerge by interacting with other entities, such as viewers or environmental factors, but also through the artwork itself as it develops over time. Daniel recounted an instance where a crack in his robotic installation continued to expand, gradually altering the artwork throughout the exhibition:

\begin{quote}
    I used [a] dome as a costume of the robot, and it (the robot) was an interactive real-time installation. The foam [on the dome] got a crack, and I decided to keep it cracking throughout the exhibition for one week. The crack in the costume was tearing down and it created a different artistic situation I could not have planned. It was so strong that it changed the whole work... I want to be sensible to those moments and see them as part of the process... I don't see that (situation) as, `okay, that is now destroying my artwork.' No, it is evolving or creating a new one within.
\end{quote}

This case illustrates how robotic artwork is not fixed but remains malleable even during the exhibition stage; temporal changes within the artwork can introduce new artistic qualities that evolve the work beyond its original design. Highlighting the artwork's temporality here allows for elucidating how the current state of the created artifact and creativity come to be. The practice of robotic art thus extends beyond the creation stage, encompassing the exhibition period. While in many cases the creation process is well planned, and temporal changes during the exhibition are typically unforeseen, both processes reveal that robotic art is in a state of ongoing creation across time. By paying attention to these processes, we unravel the temporal dimension that contributes to the creative values in robotic art.

In this Findings section, we have highlighted four aspects of robotic art practice that contribute to the artistic quality of the work or to achieving some artistic goals. The analysis reveals how various actors—artists, robots, audiences, and environments---are involved in the practice, influencing one another. These interactive patterns explain how creativity in robotic art is distributed within and emerges from the relations of actors. This idea echos with Daniel's reflection, as he noted that he sees robotic artwork as \textit{``a product of a situation of a creative potential that is part of the environment, all the entities involved as well as me,''} emphasizing the distributed and emergent nature of creativity in robotic art.

\section{Discussion}

\begin{figure*}[htbp]
    \centering
    \includegraphics[width=0.88\textwidth]{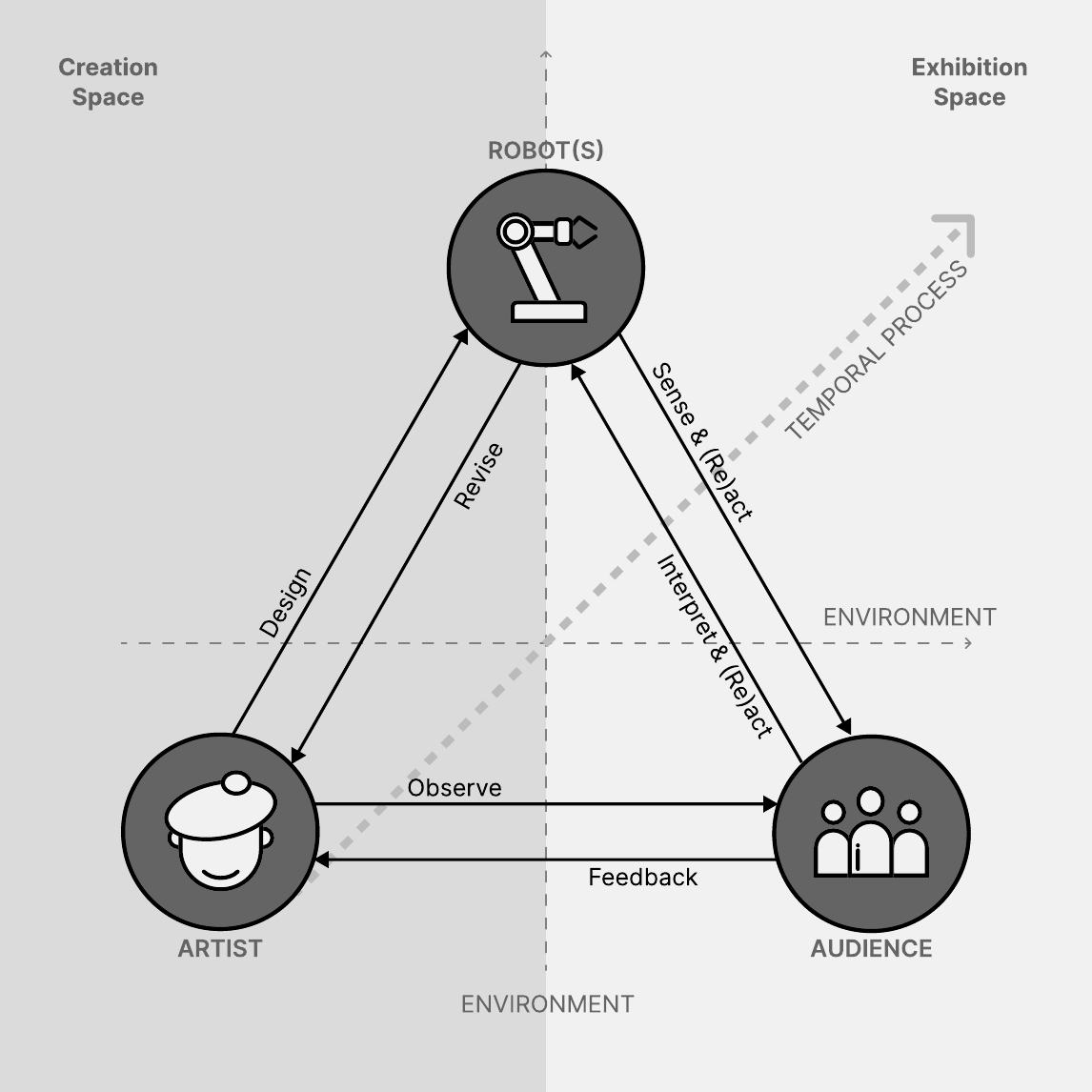}
    \caption{Actors/actants in robotic art practice and their interactive relations. Robotic art practice unfolds primarily in two spaces: the creation space where interactions happen mainly between artists and robots, and the exhibition space where interactions mostly involve audiences and robots. The two spaces constitute the ENVIRONMENT plane. Within the plane, directed arrows between the actors indicate the types of interaction. For example, the \textit{Design} arrow indicates that the artist designs the robot(s), and the \textit{Revise} arrow indicates that the robot(s) make the artist revise artistic ideas. All the actors/actants may also intra-act with the ENVIRONMENT. The actors/actants and their interactive relations may differ at different times along the axis of TEMPORAL PROCESS that is orthogonal to the plane.}
    \Description{This figure shows the actors/actants in robotic art practice and their interactive relations.}
    \label{PracticeDiagram}
\end{figure*}

This study on robotic art explores human-machine relationships in creative processes.
It first contributes as an empirical description of artistic creativity in robotic art practice---an unconventional use of robots---examined through the artists' perspectives on their creative experiences. Our analysis reveals three facets of creativity in robotic art practices: the \textit{social}, \textit{material}, and \textit{temporal}. Creativity emerges from the co-constitution between artists, robots, audience, and environment in spatial-temporal dimensions, as illustrated in \autoref{PracticeDiagram}. Acknowledging the audience as an important actor reflects the social dimension in that creativity does not stem from the artists but from their interactions with the audience. Robots are the major material and technological actants characterizing creative practices, shaping the conditions for creativity to emerge. The axis of the temporal process signifies that the practice is situated within a time continuum, and the actors/actants and their relations shift over time. In this way, temporality constitutes another dimension of creativity in robotic art.

Accordingly, as the second contribution, this study outlines implications for \textit{socially informed}, \textit{material-attentive}, and \textit{process-oriented} creation with computing systems\footnote{For the sake of clarity, we mean ``creation with computing systems'' by three types of scenarios: human creator(s) create computing system(s) as the final artifact(s) (e.g., robots are artworks themselves); human creator(s) use computing system(s) to create the artifact(s) (e.g., robots create artworks as human planned); and human creator(s) and system(s) work in tandem to produce the artifact(s) (e.g., human-robot co-creation).} to facilitate creation practices. These insights can inform related HCI research on media/art creation, crafting, digital fabrication, and tangible computing.
In each following subsection, we present each implication with a grounding in corresponding findings from our study and relevant literature in HCI and adjacent fields on art, creativity, and creation.

\subsection{Socially Informed Creation}

The sociality of creativity means that creativity is distributed among different human actors, commonly within the creators or between the creators and the audience. Glăveanu’s ethnographic study on Easter egg decoration in northern Romania~\cite{glaveanu_distributed_2014} showed that artisans anticipate how others might appreciate their work and adjust their creative decisions accordingly. Even in the absence of direct interaction, the audience’s potential responses become part of the creative process, as artisans imagine feedback and predict reactions. In this sense, the sociologist Katherine Giuffre argues that ``\textit{creative individuals are embedded within specific network contexts so that creativity itself, rather than being an individual personality characteristic is, instead, a collective phenomenon}''~\cite[p. 1]{giuffre2012collective}.

We found that the practice of robotic art manifests this sociality as it involves, particularly artists and audiences. 
Our artists take audiences' reactions to their artwork as feedback and then revise the robots' functions and aesthetics accordingly. 
For example, as shown earlier, Robert added a protective fuse onto his robot because he expected that children would squeeze the springs together and cause a short circuit; Alex's enthusiasm and attention to the audience's imagination about his robots led him to new aesthetic designs of both the robots and the scene layouts. The artists may directly ask about the audience's judgment of quality but they often receive feedback just by observing the audience's reactions or sometimes by learning from the audience's imagination about the robots.
Meanwhile, our findings reveal that audience reception is not an individual matter but is often associated with their sociocultural codes, including shared cultural norms, beliefs, expectations, and aesthetic values. The audience can be seen as representatives of these broader cultural codes. For example, Mark and Robert observed that the animist tendency in some East Asian societies is associated with higher acceptance of and interest among the audience in intelligence and agency of robots and non-human entities. Such sociocultural contexts influence not only how audiences interpret the work but also how artists anticipate and respond to these perspectives in their creative process.

A creative process, including creation and reception, is essentially a social activity. The second wave of creativity research in psychology has argued for creativity's dependency on sociocultural settings and group dynamics~\cite{sawyer2024explaining}. Recent discussions from creativity-support and social computing researchers also called for more attention to the social aspect of creativity~\cite{kato2023special, fischer2005beyond, fischer2009creativity}. There is a clear need to consider the audience when producing creative content. For instance, researchers studying video-creation support have examined audience preferences to inform system designs that align with these preferences~\cite{wang2024podreels}. Such work highlights how creative activities extend beyond individual creators to co-creators and heterogeneous audiences. Some HCI researchers conceptualize creativity as by large a socially constructed concept, perceived and determined by social groups~\cite{fischer2009creativity}. 
Prior HCI work examined the social aspects between art creators. For example, creators and performers in music and dance form social relationships through artifacts, making the final work a collaborative outcome~\cite{hsueh2019deconstructing}. There is also a system designed to support collaborative creation between artists~\cite{striner2022co}. However, the social creative process between creators and audience is less articulated in HCI. Jeon et al.'s work~\cite{jeon2019rituals} stands as an exception, suggesting that professional interactive art can involve evaluation with the audience in the creation stage. 
Another relevant approach in HCI involves enabling the general public to participate in co-creation alongside professional creators. ~\citet{matarasso2019restless}, for instance, promoted ``participatory art'' as ``\textit{the creation of an artwork by professional artists and non-professional artists working together}'' with non-professional artists referring to the general public engaged in the art-making process. Similarly, socially inclusive community-based art also considers target communities' perception of the artwork during creation~\cite{clark2016situated, clarke2014socially}. But like participatory design~\cite{schuler1993participatory}, these art projects aim for social justice more than creativity in the work~\cite{murray2024designing}, let alone that direct participation in art creation is not always feasible. Our findings suggest that feedback from the audience can lead to creative ideas, as well as that the feedback can be generative and remain low-effort for the audience.

Unlike conventional design feedback---which is typically expected to be specific, justified, and actionable~\cite{yen2024give, krishna2021ready}---the feedback that resonates with our artists is often implicit, creative, and generative. Such feedback may include audiences' imaginations stimulated by the work, personal and societal reflections, and even emotions, facial expressions, micro-actions, and observable behaviors following the art experience. Our artists gathered this implicit feedback not by posing evaluative questions, as commonly done in typical design processes (e.g., usability testing, think-aloud protocols), which seek to elicit clear, relatively structured responses. Instead, they closely observe the audience's reactions and interpret their subjective perceptions. This form of implicit feedback, while indirect, can lead to more creative ideas by embracing open, multifaceted interpretations of the work~\cite{sengers2006staying}. Computing systems for creation should better incorporate implicit feedback in addition to explicit ones from the audience into the creation process. Implicit feedback can be indirect, creative, inspirational, and heuristic about functions and aesthetics. A hypothetical instance of such design can be a system that helps creators perceive audiences' implicit reactions and perceptions and variously interpret them, for further iteration.

Moreover, as seen in Robert and Daniel's experiences, the audience may participate in robotic live performances by interacting with the robots, who may change actions accordingly, triggering a loop of simultaneous mutual influence that makes the work performative and improvisational.
HCI researchers explored performative and improvisational creation with machines, focusing on developing and evaluating systems with performative capabilities, including music improvisation with robots~\cite{hoffman2010shimon}, dance with virtual agents~\cite{jacob2015viewpoints, triebus2023precious}, and narrative theatre~\cite{magerko2011employing, piplica2012full}. \citet{kang2018intermodulation} discussed the improvisational nature of interactions between humans and computers and argued that an HCI researcher-designers' improvisation with the environment facilitates the emergence of creativity and knowledge. Designs of computing systems for creation can leverage performativity in service of creative experience. One possible direction could be to allow the audience to embed themselves in and interact with elements of static artwork in a virtual space, turning the exhibition into an improvisational on-site creation~\cite{zhou2023painterly}.
While interactions with machines during performance are mostly physical or embodied, we posit that they can also be a \textit{symbolic engagement}. Alex's audience projected themselves and their personalities onto his robots, which established a symbolic relevance, generating creative imaginations. During exhibitions, East Asian audiences carried the animist views shaped by their sociocultural backgrounds, and robots, through the performance, were successful in symbolically matching the views, stimulating aesthetic satisfaction. Symbolic engagement resonates with what ~\citet{nam2014interactive} called the ``reference'' of the interactive installation performance to participants' sociocultural conditions.
As such, we propose that designers of computing systems for creation may consider establishing symbolic engagement between the produced artifacts and the audience as a way to enhance perceived creativity or enrich the creative experience. One example is an interactive installation, \textit{Boundary Functions}~\cite{snibbe1998}, which encourages viewers to reflect on their personal spaces while interacting with the installation and others. Another example is \textit{Blendie}, a voice-controlled blender that requires a user to ``speak'' the machine's language to use it. This interaction builds a symbolic connection between the user and the device, transforming the act of blending into a novel experience~\cite{dobson2004blendie}.

\subsection{Material-Attentive Creation}

The theory of distributed creativity by Glaveanu claims that creativity distributes across humans and materials, so the creation practice itself is inevitably shaped by objects~\cite{glaveanu_distributed_2014}. In his case of Easter egg decoration, materials are not passive objects but active participants in artistic creation; e.g., the egg decorators face challenges from color pigments not matching the shell, wax not melted at the desired temperature, to eggs that break at the last step of decoration; hence, materials often go against the decorators' intentions and influence future creative pathways~\cite{glaveanu_distributed_2014}.
Materials manifest specific properties, which afford certain uses of the materials while constraining others~\cite{leonardi2012materiality}. Our findings highlight the critical role of materiality in artistic practice, showing that artists intentionally arrange materials to enhance the creative values of their work.

Robotic art relies on the material properties of robots and other objects. An apparent property of most materials is their physicality~\cite{leonardi2012materiality}, meaning they possess a tangible presence that enables interaction with other physical entities. Here, we consider physicality and embodiment interchangeable as computational creativity researchers have conceptualized~\cite{guckelsberger2021embodiment}.
Our findings support both the conceptual and operational contributions of embodiment for creative activities. For the conceptual aspect, the embodied presence of robotic systems supports creative thinking for our artists, exemplary in Linda's case where she found new art ideas around the difference between human and robot bodies through bodily engagement with robots. 
For the operational aspect, the embodied nature of robotic artworks and their creation processes exhibit original aesthetics that are based on physics much different from disembodied works, e.g., embodied drawings by David's non-industrial robotic arms are dynamic due to physical movements and thus artistically pleasant, which is hard to replicate in simulated programs.

These findings on embodiment of robotic art (Section \ref{f:emb}) closely relate to HCI's attention on embodied interaction as a way to leverage human bodies and environmental objects to expand disembodied user experiences. 
For example, as~\citet{hollan2000distributed} explained, a blind person's cane and a cell biologist's microscope as embodied materials are part of the distributed system of cognitive control, showing that cognition is distributed and embodied. 
Similarly, theories of embodied interaction in HCI explicate how bodily interactions shape perception, experience, and cognition~\cite{marshall2013introduction, antle2011workshop, antle2009body}, backed up by the framework of 4E cognition (embodied, embedded, enactive, and extended)~\cite{wheeler2005reconstructing, newen20184E}. 
Prior works suggest that creative activities with interactive machines rely on similar embodied cognitive mechanisms ~\cite{guckelsberger2021embodiment, malinin2019radical}, which are operationalized by tangible computing~\cite{hornecker2011role}. 
As related to robots in creation, HCI researchers show that physicality or embodiment of robots in creation may lead to some beneficial outcomes, such as curiosity from the audience, feelings of co-presence, body engagement, and mutuality, which are hard to simulate through computer programs~\cite{dell2022ah, hoggenmueller2020woodie}. Embodied robotic motions convey emotional expressions and social cues that potentially enrich and facilitate creation activities like drawings~\cite{ariccia2022make, grinberg2023implicit, dietz2017human, santos2021motions}. Guckelsberger et al.~\cite{guckelsberger2021embodiment} showed in their review that embodiment-related constraints (e.g., the physical limitations of a moving robotic arm) can also stimulate creativity. These constraints push creators to develop new and useful movements, echoing the broader principle that encountering obstacles in forms or materials can lead to generative processes. This phenomenon is similarly observed in activities such as art and digital fabrication~\cite{devendorf2015being, hirsch2023nothing}. In co-drawing with robots, physical touch and textures of drawing materials made the artists prefer tangible mediums (e.g., pencils) than digital tools (e.g., tablets) that fall short in these respects~\cite{jansen2021exploring}.

Materiality plays a crucial role in the embodiment of robots, as the choice of materials fundamentally shapes the physical forms and properties. This focus on materials extends to art practices, where robots made with soft materials introduce new aesthetics and sensory experiences~\cite{jorgensen2019constructing, belling2021rhythm}, and the use of plants and soil in robotic printing creates unique visual effects~\cite{harmon2022living}. Following Leonardi's ~\cite{leonardi2012materiality} conceptualization of materiality, we refer to the materials of robots as encompassing physical and digital components---including the shell, hardware, mechanical parts, software, programs, data, and controllers---each significant to the artist's intent. ~\citet{nam2023dreams} found that the material constraints of robots can limit creative expression but simultaneously stimulate creativity when artists push the boundaries.


Even carefully designed, digital and mechanical components in robots are prone to errors or bugs in everyday runs, causing malfunctions or unexpected consequences. This reflects the unique materiality of robots as complex computing systems. From an engineering perspective, errors signal unreliability and must be eliminated, driving advancements in robotics---where error detection and recovery are central~\cite{gini1987monitoring}---as well as in digital fabrication, which prioritizes precision over creative exploration~\cite{yildirim2020digital}. 
However, material failures and accidents are inevitable, exemplifying what has been called the `craftsmanship of risk'~\cite{glaveanu_distributed_2014} in material art. For our artists, these risks are often creatively utilized and incorporated into their work: these moments of breakdown---whether physical or digital---become resources for new creative expression. Errors are anticipated and intentionally designed into the process and work of our artists. In some cases, such as for Alex, the entire concept of one of his works is machine errors.

Reports on how artists view errors within engineering and creation processes are dispersed throughout HCI literature. ~\citet{nam2023dreams} showed that the accumulation of ``contingency'' and ``accidents''---unexpected, serendipitous, and emergent events during art creation like errors---meaningfully constituted the final presentation of the artwork. Song and Paulos's concept of ``unmaking'' highlighted the values of material failures in enabling new aesthetics and creativity~\cite{song2021unmaking}. Kang et al.~\cite{kang2022electronicists, kang2023lady} introduced the notion of an ``error-engaged studio'' for design research in which errors in creative processes are identified, accommodated, and leveraged for their creative potential. Collectively, these works advocate for reframing errors from something to avoid to something to embrace and recognize. We want to push this further by arguing that errors can be intended and be part or sometimes entire of the design. Several artists, including participants from our study, have been deliberately seeking errors to formulate their designs. Roboticist Damith Herath recounted when he mistakenly programmed a motion sequence of a robotic arm, his collaborator, robotic artist Stelac responded with ``[W]e need to make more mistakes;'' as many mistakes were made, the initial pointless movements became beautiful, rendering the robot ``alive'' and ``seductive'' \cite{herath2016robots}. Similarly, AI artists sometimes look for program glitches to generate unusual styles and content~\cite{chang2023prompt}. Therefore, creators may not only passively accept errors but can actively seek and utilize them. Errors can be integral to the design itself---errors can \textit{be designed into} an artifact, and the design/idea of the artifact can be all about errors.

Thus, to focus on material-attentive creation---considering the creative arrangement of materials---we suggest exploring the embodiment and materiality of creation materials, objects, and environments to recognize their creative potential. 
Specifically, we propose using a design method/probe that enables creators to realize both the conceptual and operational contributions of materiality. This approach may build on the material probe developed by~\citet{jung2010material}, which calls for exploring the materiality of digital artifacts. A material-attentive probe would enable creators to engage with diverse materials, objects, and environments through embodied interaction, encouraging them to speculate on material preferences and limitations, and to compare and contrast material qualities---insights that can inform creative decisions.
To accommodate, seek, and actively harness the creative potential of errors, we propose embracing failures, glitches, randomness, and malfunctions in computing systems as critical design materials---elements that creators can intentionally control and manipulate. By doing so, we can begin to systematically approach errors. For instance, as part of the design process, we may document how to replicate these errors and changes, allowing creators to explore them further at their discretion. This could include intentionally inducing errors or random changes to influence the creative process or outcomes.

\subsection{Process-Oriented Creation}

As shown in our findings, the creation process itself embeds creative values and meanings, and experiencing the process can be pursued as the goal of creation with computing systems.
For the robotic artists in our study, artistic values were often placed on the creation process rather than the outcome.  For example, in Alex's robotic live drawing performance, the drawing process is more important than the drawn pattern on canvas. Techniques used, decisions made, or stimuli received by robots during creation or exhibition reflect artistic ideas and nuanced thinking, as seen in Sophie's exploration of interactive decision-making in robotic drawing.

Previous HCI work has touched on the value of the process of creation. ~\citet{bremers2024designing} shared a vignette where a robotic pen plotter simultaneously imitates the creator's drawing, serving as a material presence rather than a pragmatic co-creator; here the focus of the work is no longer the outcome but the process of drawing itself. ~\citet{devendorf2015reimagining} concluded that performative actions of digital fabrication systems, rather than the fabricated products themselves, convey artistic meanings tied to histories, public spaces, time, environments, audiences, and gestures. This emphasis on process is particularly significant for media such as improvisational theatre, where the creation itself is an integral part of the final work~\cite{o2011knowledge}. ~\citet{davis2016empirically} named their improvisational co-drawing robotic agents as ``casual creators,'' who are meant to creatively engage users and provide enjoyable creative experiences rather than necessarily helping users make a higher quality product. Shifting the focus from product to process and experiences \textit{in} creation may generate alternative creative meanings.

Our artists pointed out that even a ``finished'' artwork in an exhibition is not truly finished. A crack in Daniel's robotic artwork introduced a new artistic meaning, ultimately subverting the entire work. As the properties of the work change over time---whether due to the artist's intent, material characteristics, or environmental factors---the artwork evolves, revealing new aesthetics and meanings. 
Based on these observations, we argue that creation processes should not be regarded as one-shot transactions, as creative artifacts, particularly physical ones, continue to change and generate artistic values. For instance, material wear and destruction bring unique aesthetics, often contrasting with the original form ~\cite{zoran2013hybrid}, and are seen as signs of mature use~\cite{giaccardi2014growing}.
Changes such as material failure, destruction, decay, and deformation---what~\citet{song2021unmaking} referred to as ``unmaking,'' a process that occurs after making---meaningfully transforms the original objects. Similarly, through Broken Probes, a process of assembling fractured objects, ~\citet{ikemiya2014broken} demonstrated that personal connections, reminiscence, and reflections related to material wear and breakage add new values to the objects. Drawing from Japanese philosophy Wabi-Sabi, ~\citet{tsaknaki2016expanding} reflected on the creeds of `Nothing lasts,' `Nothing is finished,' and `Nothing is perfect' and pointed to the impermanence, incompleteness, and imperfection of artifacts as a resource that designers, producers, and users can utilize to achieve long-term, improving, and richer interactive experience~\cite{tsaknaki2016things}. Insights from this study contribute to this line of thought by showing how robotic artists appreciate the aesthetics and meanings of temporal changes after the creation phase.

The findings underscore the need to reconceptualize creation as encompassing more than just the process aimed at producing a final product; it also includes what we term \textit{post-creation}. Distinct from repair, maintenance, or recycle, \textit{post-creation} entails anticipating and managing how an artifact evolves after its ``completion'' in the conventional sense. Specifically, we encourage creators to anticipate and strategically engage with the post-creation phase, considering potential changes to the artifact and their consequences for interactions with human users. For instance, during the creation process, creators may focus on possible material changes the artifact might undergo post-creation, allowing them to either mitigate or creatively exploit these potential changes. This expanded view of creation invites us to trace post-creation developments and to plan how our creative intentions can be embedded in its potential degradation, transformation, or evolution over time.

We categorize the design implications into three aspects, but we do not suggest that a computing system must implement all simultaneously, nor that each aspect should be considered in isolation. Social interactions, such as those between artists and audiences, already presume the presence of material actants like robots, and these interactions inform future arrangements of materials. Thus the social and material aspects can be entangled and mutually constitutive as seen in sociomaterial practices~\cite{orlikowski2007sociomaterial, cheatle2015digital, rosner2012material}. The temporal aspect is orthogonal to the other aspects because both social interactions and material manifestations unfold and shift in a temporal continuum.

\section{Conclusion}

This study empirically examines the unique practice of using robots in artistic production. We uncover the social, material, and temporal dimensions of this practice that shape perceptions of creativity in robotic art. Our analysis highlights the role robots play in driving artistic inquiry and fostering creative thinking, expression, and experimentation. Insights from robotic art offer implications for understanding how creativity emerges through interactions among humans, technological systems, and material actants over time. These findings also suggest ways to facilitate creative activities through deeper engagement with computing systems. As a first step, we propose a framework to advance creative practices with computing systems by promoting \textit{socially informed}, \textit{material-attentive}, and \textit{process-oriented} designs that better support human creators.

\begin{acks}
We are in debt to all robotic artists who participated in this study. We also thank the anonymous reviewers for their feedback.
\end{acks}

\bibliographystyle{ACM-Reference-Format}
\bibliography{main}

\appendix

\section{Coding Categories and Descriptions}
\label{app: coding}
\begin{table}[hbp]
\centering
\begin{tabular}{p{0.24\linewidth} p{0.1\linewidth} p{0.539\linewidth}}
\toprule
Category & Code Count & Description\\
\midrule
MOTIVATION & 10 & Motivations for doing robotic art \\
IDEATION & 5 & Generation of artistic ideas\\
CREATIVITY & 9 & Forms and manifestations of creativity of robotic art \\
MATERIALITY& 10 & Embodied and material aspects of robotic art \\
RELATIONSHIP & 18 & Relationships between artists and robots \\
VALUE & 11 & Important artistic values \\
BENEFIT & 3 & Benefits of doing robotic art \\
CHALLENGE & 9 & Challenges of doing robotic art \\
MISC. & 26 & Uncategorized open codes and index codes\\
\bottomrule
\end{tabular}
\caption{Coding categories and descriptions}
\label{tab:categories}
\end{table}

\end{document}